\documentstyle[12pt,hpatex]{article}  
\def\beq{\begin{equation}}
\def\eeq{\end{equation}}
\def\d{\delta}
\begin{document}
\title{CMBR dipole from ultra large scale isocurvature perturbations}
\authors{David Langlois}
\address{ D\'epartement d'Astrophysique Relativiste et de Cosmologie,\\
Centre National de la Recherche Scientifique,\\ Observatoire de
Paris, 92195 Meudon, France}
\abstract{The observed CMBR dipole, generally interpreted as the consequence 
of the peculiar motion of the Sun with respect to the reference frame
of the CMBR, can also be explained by the presence of
 ultra large scale (of the order of $100H_0^{-1}$) 
 isocurvature perturbations. 
Moreover, the simplest model of inflation with several scalar fields, 
namely that of double inflation, with appropriate parameters, can produce 
 such perturbations. }

\section{Introduction}
The dipole moment of
the  Cosmic Microwave Background Radiation (CMBR) anisotropy 
 is larger than the quadrupole, measured recently by the satellite
COBE,  
by two orders of magnitude (\cite{kogutetal}). In general, this dipole 
is interpreted 
as the peculiar motion of the Sun with respect to the CMBR ``rest 
frame'', whereas the other multipoles are seen as a consequence of 
primordial cosmological perturbations of a homogeneous and isotropic 
universe. 

However, from a theoretical point of view,  the dipole or part of it
 can also be of cosmological origin (\cite{ppt}, \cite{lp}). 
From an observational point of view, although there is a general trend
to favour the Doppler effect,  this question is not yet settled.
This contribution  summarizes the conclusions of 
two recent works (\cite{lp},\cite{l96}) on an alternative explanation for the 
dipole: superhorizon isocurvature perturbations.

One must be aware that  the origin of the dipole affects the interpretation
of the measured quadrupole. Indeed, the Doppler effect, whose
complete expression is 
\beq
T_{obs}({\vec e})=T(1-v^2/c^2)^{1/2} (1-{\vec e}.{\vec v}/c)^{-1},
\eeq
 ($\vec e$ gives the direction of observation on the celestial sphere)
 also induces a quadrupole, which must be substracted from the observed dipole
to obtain the ``true cosmological quadrupole''. The latter would be modified 
if the dipole turns out  not to be a Doppler dipole. 

\section{Ultra large scale isocurvature perturbations}

 The  CMBR fluctuations at large angular scales can be written as 
the sum of two contributions: the intrinsic fluctuations on the 
last scattering surface; the Sachs-Wolfe fluctuations due to the 
presence of geometrical perturbations on the light trajectories.
In a ``scalarly'' perturbed flat FLRW model, with the metric
\beq
ds^2=-(1+2\Phi)dt^2+a^2(t)(1-2\Phi)\delta_{ij}dx^idx^j,
\eeq
the Sachs-Wolfe contribution is simply
\begin{equation}
\left({\Delta T\over T}\right)_{SW}(\vec e)={1\over 3}\left[\Phi_{ls}-\Phi_0
\right]-\vec e.\left[\vec v_{ls}-\vec v_0\right], \label {sw}
\end{equation} 
 where the subscript $ls$ refers to the last
scattering surface and the subscript $0$ to the observer today. 
In a  universe with  
radiative matter (with a density contrast $\d_r\equiv \d\rho_r/\rho_r$)
and  pressureless matter (with a density contrast 
$\d_m\equiv \d\rho_m/\rho_m$), matter perturbations can be decomposed 
into  adiabatic perturbations, for which the intrinsic contribution is 
negligible, and  isocurvature perturbations, for which 
\beq
\left({\Delta T\over T}\right)_{int}\simeq -{1\over 3}S,
\eeq
where $S\equiv \d_m-(3/4)\d_r$ is the entropy perturbation.

The observed CMBR temperature fluctuations are generally decomposed in
spherical harmonics : 
\begin{equation} 
{\Delta T\over T}(\theta, \phi)=
\sum_{l=1}^{\infty}\sum_{m=-l}^{l} a_{lm}Y_{lm}(\theta,\phi). 
\end{equation}
Assuming that the cosmological perturbations can be described as a 
homogeneous random field, the power spectrum being defined by
$\langle \Phi_{\bf k} \Phi_{\bf k}'\rangle=2\pi^2k^{-3}{\cal P}_\Phi(k)\delta(
{\bf k}-{\bf k}')$ ($\Phi_{\bf k}=(2\pi)^{-3/2}\int d^3{\bf x}\Phi({\bf x})
e^{-i{\bf k.x}}$),
the prediction for the harmonic components  
$l\ge 2$  on large angular scales, 
in the case of adiabatic perturbations, is
\begin{equation}
\Sigma_l^2=\langle|a_{lm}|^2\rangle={4\pi\over 9} \int {dk\over k} 
{\cal P}_\Phi(k)  j_l^2(2k/a_0H_0).  \label{sigma}
\end{equation}
As  shown in \cite{lp}, if our peculiar velocity is ignored, the dipole
from adiabatic perturbations can never be bigger than the quadrupole. This
is due to a cancellation between  the terms $\Phi /3$ and ${\bf e}.{\bf v}$ 
in the dipole. However, in the case of isocurvature perturbations, because
of the large intrinsic contribution, the dipole can be 
made much larger than the quadrupole by considering perturbations larger
than the present Hubble  radius. The resulting ``cosmological'' dipole
will then be \cite{lp}
\begin{equation}
 a_1^2\equiv{3\over 4\pi}\Sigma_1^2={1\over 3} \int {dk\over k} 
{\cal P}_S(k)  j_1^2(2k/a_0H_0).  \label{dipole}
\end{equation}
The question now arises what could be the physical origin of such 
perturbations. The next section provides a possible answer, in the framework
of double inflation.

\section{ The double inflation model}

The simplest model of double inflation (see e.g. \cite{ps1}), consisting of two
minimally coupled massive scalar fields, is  described by the Lagragian
\begin{equation} 
{\cal L}={{}^{(4)}
R\over 16\pi G}-{1\over 2} \partial_\mu\phi_l\partial^\mu\phi_l
 -{1\over 2} m_l^2\phi_l^2 -{1\over 2} \partial_\mu\phi_h\partial^\mu\phi_h
 -{1\over 2} m_h^2\phi_h^2,
\end{equation}
where $\phi_l$ and $\phi_h$ are respectively the light and heavy scalar
fields; ${}^{(4)}R$ is the scalar spacetime curvature and $G$ is 
Newton's constant. In addition to  $m_l$ and $m_h$,
which from now on are assumed to satisfy  $m_h/m_l \gg 1$, there
is a third parameter $s_0$ corresponding to the initial conditions for 
the two scalar fields ($s_0=4\pi G\phi_h^2$ when $\phi_h=\phi_l$).
It is convenient to label any moment during inflation by the corresponding
number of e-folds before the end of inflation, s. There is a natural scale
associated with it, which is the comoving Hubble radius at this moment.

The first phase of inflation, driven   by the heavy scalar field,
lasts  until equality of the 
energies of the two scalar fields ($m_h\phi_h=m_l\phi_l$), which  
occurs for  $s_b\simeq s_0$.
If  $s_0>(m_h/ m_l)^2$, which will be assumed here,  inflation goes on,
 now driven by the light scalar field.
Meanwhile the heavy scalar field  continues its slow-rolling until
 $H\sim m_h$, corresponding to $s=s_c$,
after which it oscillates.
Finally  the Hubble parameter reaches the value $m_l$ 
and inflation stops. 
In the ``standard''  model of double inflation, the scales $\lambda_b$
and $\lambda_c$ (one always has $\lambda_b>\lambda_c$)  are within the 
Hubble radius today, denoted $\lambda_H$, (\cite{ps1}) 
whereas, here, the parameters 
of the model are chosen such that these two scales are {\it outside} 
the present Hubble radius $\lambda_H$.

The spectrum of adiabatic perturbations today, which it is convenient 
to express in terms of $\Phi$, is computed from the vacuum 
quantum fluctuations of the two scalar fields in the inflationary period.
This spectrum has two plateaus, as explained in \cite{ps1}. 
Since, here, the transition scale is larger than the Hubble radius 
today, only the lower plateau, corresponding to the second phase of 
inflation (driven by the light scalar field), and given by
\begin{equation}
{\cal P}_\Phi={12\over 25\pi}s^2{m_l^2\over m_P^2},
\end{equation} 
will be accessible to observations ($s(k)\simeq \ln (k_e/k)$).
Assuming the quadrupole  to be essentially due to adiabatic perturbations.
 the COBE measurement
 of $Q_{rms-PS}(n=1)\simeq 18 \mu K$ (\cite{banday}) implies 
$m_l\simeq  7.8\times 10^{-5} m_P s_H$ (for $s_H=60$, 
 $m_l\simeq 1.3 \times 10^{-6} m_P$).

In addition to adiabatic pertubations, this model can also produce 
 isocurvature perturbations. This is the case for example if one assumes
that the light scalar field, after inflation, will decay into ordinary 
matter, while the heavy scalar field remains decoupled from ordinary 
matter.
The subsequent spectrum of isocurvature perturbations is given by
(\cite{l96}, \cite{ps2})
\begin{equation} 
{\cal P}_S={4 G\over 3\pi}m_l^2\left({s_0\over s}\right)^{m_h^2/m_l^2}
\end{equation}
until  $s=s_c$, after which  the production of isocurvature perturbations
falls abruptly.
Substituting this spectrum in (\ref{dipole}) and introducing the variable 
$X=\ln (k_c/ k)$, the expected dipole reads
\begin{equation}
a_1^2= {16\over 81\pi}\left({m_l\over m_P}\right)^2 e^{2(s_H-s_c)}
\left({s_0\over s_c}\right)^{m_h^2/m_l^2}
\int_0^\infty dX {e^{-2X}\over (1+X/s_c)^{m_h^2/m_l^2}}, \label{a1}
\end{equation}
 where has been used the fact that $j_1(x)\sim x/3$ for small $x$.
The integral in (\ref{a1}) 
is well approximated as $(2+(m_h/m_l)^2/s_c)^{-1}$.
For consistency, one must also check that the quadrupole 
generated by isocurvature perturbations
is smaller than the adiabatic quadrupole.
This requires, as seen in \cite{lp}, that 
\begin{equation}
s_c >  s_H + 4.6, \label{cond}
\end{equation}
i.e. that the cut-off scale $\lambda_c$ is one hundred times larger than
the Hubble radius today.
The value $s_H$ is of the order of $60$, its exact value 
depending  on the  history of the universe. 
 Because of the power law dependence
in (\ref{a1}), {\it  the dipole can be  much bigger 
than the quadrupole}. 
 Because of the condition (\ref{cond}), implying that $m_h^2/m_l^2$
must be large, the main constraint, surprisingly,  comes
from requiring that the dipole is not too high with respect to the 
quadrupole: this requires that   $s_0/s_c$ must be  very close to 1.
In this range of parameters, double inflation thus provides adequate spectra
of perturbations, explaining the observed dipole and higher multipoles
without a Doppler effect.

%*******************************************************************
 
\end{document}